# High intrinsic flux pinning strength of BaFe$_{2-x}$Co$_x$As$_2$ superconductor

K. Vinod, Shilpam Sharma, A. T. Satya, C. S. Sundar and A. Bharathi*

Condensed Matter Physics Division, Materials Science Group,

Indira Gandhi Centre for Atomic Research, Kalpakkam, India - 603102

**Abstract**

The field dependence of flux pinning potential, $U_0$, for under-doped, optimally doped and over-doped Ba(Fe,Co)$_2$As$_2$ single crystals has been investigated. $U_0$ is determined using the Thermally Activated Flux Flow (TAFF) model, for both H || ab and H || c configurations. The observed power law dependencies of $U_0$ versus H suggest that for fields upto 12 T for H || ab and 7 T for H || c, individual flux pinning, and for fields beyond 7 T for H || c, collective pinning becomes operative. A comparison of $U_0$ for the Ba(Fe,Co)$_2$As$_2$ system with some other superconductors is presented.

*Corresponding author

A. Bharathi

Condensed Matter Physics Division,Materials Science Group,

Indira Gandhi Center for Atomic Research,

Kalpakkam, 603102 India.Tel: 91 44 27480081, Fax: 91 44 27480081

E-mail: bharathi@igcar.gov.in

High transition temperature ($T_C$), large upper critical field ($H_{C2}$), high irreversibility fields ($H_{irr}$), large critical current density ($J_C$), low anisotropy and good inter-granular connectivity are essential characteristics of superconducting materials for magnet applications. The two layer Ba arsenide (Ba122) superconductors, whose fundamental properties have been extensively investigated due to the availability of good quality single crystals, seem to possess large $H_{C2}$, $H_{irr}$ and $J_C$ along with low anisotropy.[1-4] The high $J_C$ of $10^5$-$10^6$ A/cm$^2$ [2,3] in seemingly defect free single crystals of doped BaFe$_2$As$_2$ (Ba122) is surprising and elicits questions on the origin of flux pinning centres in these materials. To answer this, real space visualization of the flux line lattice has been carried out using magneto-optical imaging, scanning tunneling spectroscopy and scanning SQUID microscopy,[5-7] all of which show the presence of a disordered vortex lattice, suggesting strong bulk pinning of vortices in these materials.[5-7] Based on polarized-light imaging, x-ray diffraction and magnetic measurements, it was argued[3] that the presence of twin boundaries enhance flux pinning in the optimally doped sample giving rise to large $J_C$. In agreement with these findings, SQUID microscopy images of the vortex lines suggest that twin boundaries act as barriers for the motion of the flux lines.[8] This however is inconsistent with the observation of large $J_C$ in the overdoped regime where twinning is absent.[7] In addition, since coherence length in the Ba122 arsenides is about few nanometers, the efficacy of micrometer sized twin boundaries acting as good flux pinning centres seems questionable.

Apart from direct imaging techniques, study of flux dynamics viz., the study of the variation in the magnitude flux pinning potential ($U_0$), under magnetic field can provide valuable information on the nature of flux pinning. In addition, the magnitude of $U_0$ is an important factor that can affect $J_C$ and Hirr and in turn determine the technological worth of a superconducting material. In the weak pinning regime, phenomenological theories [9,10] exist that predict that the activation energy to move a bundle of a two dimensional array of rigid fluxons is inversely proportional to the linear power of flux density ($U_0 \propto \phi_0/B$). This gets modified in the presence of thermal vibration of each fluxon, resulting in $U_0 \propto \phi_0/B^{0.5}$.[9,10] The presence of inter fluxon interaction is seen to reduce the exponent even further.[10] In the opposite limit of flux pinning at individual fluxons, $U_0$ depends weakly on the magnetic field.[11] Experimentally, few investigations of the variation of $U_0$ with magnetic field have been carried out in single crystals of arsenide superconductors.[12,13] Studies in $(Ba,K)Fe_2As_2$ single crystals in fields upto 13 T, indicate a large field independent $U_0$,[12] whereas similar work carried out in superconducting oxy-arsenide single crystals indicate a smaller $U_0$, whose field dependence changes at 3 T, suggestive of a changeover in the nature of pinning.[13] In the $Ba(Fe,Co)_2As_2$ system the magnitude of flux pinning potential and its variation with magnetic field has not yet been investigated systematically, although the search for pinning centres responsible for flux pinning and its correlation to high $J_C$ has been examined.[3,8] Here we report on measurements of pinning potentials and their magnetic field dependencies for three different concentrations in the superconducting single crystals of $BaFe_{2-x}Co_xAs_2$, to highlight the effects due to disorder. The Co concentrations belong to the under-doped (x=0.082), close to optimum doped (x=0.117) and over-doped (x=0.143) parts of the phase diagram. The $BaFe_{2-x}Co_xAs_2$ superconducting single crystals were prepared by slow cooling without any flux as described in ref [14].

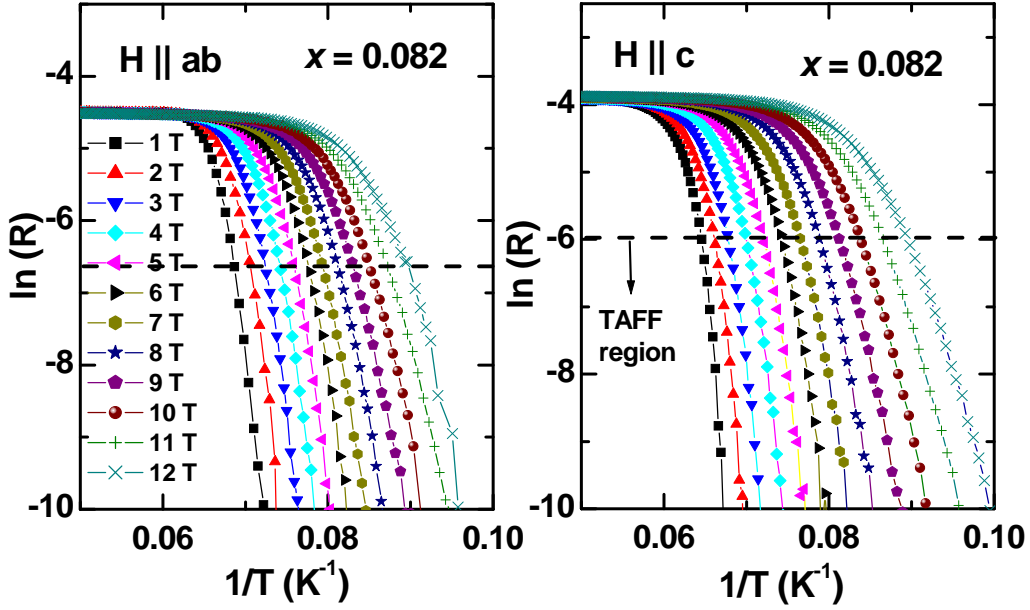

FIG 1: (Color online) Arrhenius plot for the $BaFe_{2-x}Co_xAs_2$ (x = 0.082) crystal for H || ab and H || c directions. TAFF region occurs below the dashed line. The slope of the curve in this region is used to determine $U_0$.

In the thermally activated flux flow (TAFF) model, the temperature and field dependent resistance is described by $R(T,H) = R_0 \exp[-U_0/k_B T]$, where $R_0$ is a fitting parameter, $k_B$ is the Boltzmann constant and $U_0$ is activation energy (pinning potential) for vortex bundle hopping.[15] Our measurements employ a ~2ma current on crystals with typical dimension of ~ 1 mm x 1 mm x 0.2 mm. This implies that the transport current density is much smaller than $J_C k_B T/U_0$, ensuring the necessary condition for the applicability of the TAFF model.[15,16] $U_0$ versus magnetic field, parallel to the ab plane (H || ab) and parallel to the c axis (H || c) were obtained for two set of crystals (S1 and S2), for each Co composition. Fig.1 shows ln (R) vs 1/T for one of the samples with Co fraction of x = 0.082, measured with H || ab and H || c directions. The observation of linear dependence close to the downset of the superconducting transitions, below the

dashed line indicated in Fig.1, viz., for $R(T) < 0.1R_n$, where $R_n$ is the normal state resistance, indicates that TAFF dominates the resistivity behaviour at these temperatures. From measurements of the slope of each of these curves [$U_0 = -d(\ln R)/d(1/T)$], $U_0$ at the particular applied field is obtained.

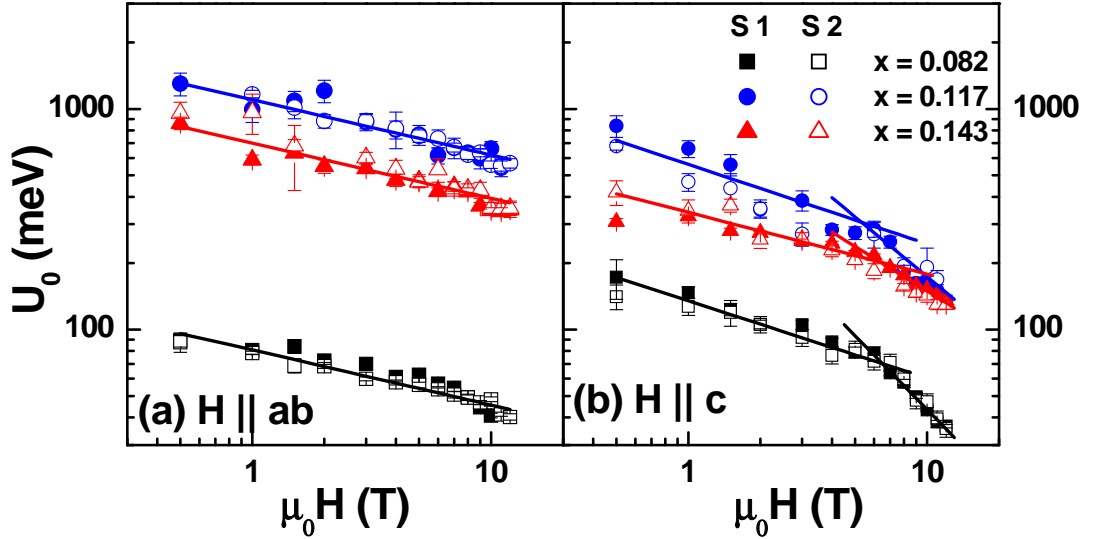

FIG 2 : (Color online) Pinning potential, $U_0$ obtained from R(T,H), for the x = 0.082, x = 0.117 and x = 0.143 $BaFe_{2-x}Co_xAs_2$ samples for (a) H || ab and (b) H ||c directions. The open and filled symbols correspond to measurements on two crystals (S1 & S2), for each composition. Solid lines shows fit with $U_0 \propto H^{-n}$, and the fit shows change in the exponent 'n' at higher fields for the H || c direction.

The $U_0$ values in the three Co substituted samples, for H || ab and H ||c, are compared in Fig. 2. The agreement of the $U_0$ in two sets of samples for each composition testifies the reliability of the data. It is clear from Fig. 2 that $U_0$ is the highest for the optimally doped sample. Further, it is notable that for the optimally and over doped Co concentrations the value of $U_0$ is higher for the H || ab direction as compared to the H || c

direction. Whereas, for the under doped composition, $U_0$ for H || c direction is slightly higher compared to that in the H || ab direction at lower fields. The larger value of $U_0$ for H || ab compared to that for H || c observed in the optimum and overdoped samples, could arise on account of the larger $H_{C2}$ along the H || ab direction (since $U_0$ depends on the superconducting condensation energy.[9])  The variation in the magnitude of $U_0$ versus Co concentrations at a given field shown in Fig. 2 seems to be consistent with the $J_C$ variation observed with Co concentration.[3]

Table 1 : Value of the exponent 'n' for $U_0 \propto H^{-n}$, for the three Co , for H || ab and H || c directions.

|  | Value of the exponent 'n' | | |
| --- | --- | --- | --- |
| Sample | H || ab | H || c | |
|  |  | H < 7 T | H > 7 T |
| x = 0.082 | 0.22 ± 0.02 | 0.34 ± 0.04 | 1.19 ± 0.07 |
| x = 0.117 | 0.27 ± 0.02 | 0.36 ± 0.04 | 0.88 ± 0.19 |
| x = 0.143 | 0.28 ± 0.03 | 0.29 ± 0.03 | 0.65 ± 0.09 |

It is evident from the Fig. 2 that $U_0$ fits to a single power law for H || ab and is proportional to $H^{-n}$ with n ~ 0.25 (see table.1).  However, for the H || c direction $U_0$ shows a faster variation with H, with different power-law behaviours, in two field ranges. The exponent for the H || c, is n=0.29-0.36 at lower fields and changes to n=0.65-1.19 beyond ~ 7 T, for all the three Co concentrations (see table.1). It is clear from table.1 that for H || c and for magnetic fields >7 T, the field exponent is close to 1 for the lowest concentration of Co. The observation of an exponent of n~1 implies that weak pinning of

flux bundles [9,10] dominates the magnetic field dependence for H || c, for the lowest Co concentration. The exponent in this field regime shows a clear decrease with increase in Co concentration, viz., the high field exponent reduces from 1 to 0.65. Since more disorder is introduced due to Co addition, it is likely that flux pinning from individual pinning centres starts competing with bulk pinning, leading to the reduction of the exponent. For H || c upto 7 T, the field exponent of $U_0$ is even more reduced with n ~ 0.33-0.29, as compared to that in fields >7 T, implying again that individual pinning from disorder dominates in this regime. This is understandable, since inter-fluxon interactions would reduce with decrease in flux density at lower fields and individual pinning of flux lines would start to dominate. Along the H || ab direction the field exponent is ~0.25 upto fields of 12 T, which again is closer to the exponent expected for individual flux pinning rather than for bulk pinning.

The results of table.1 suggest that the field exponent of $U_0$ depends on the interplay between robustness of superconducting order, individual flux pinning strength and bulk flux pinning strength. At low flux density and in the presence of a robust superconducting order, individual flux pinning dominates and $U_0$ is weakly dependent on H. [9,10] At higher field density inter fluxon interactions start playing a role and $U_0$ is determined by collective behavior of fluxons and the field exponent becomes larger although, moderated by the degree of disorder. This is brought out nicely from the $U_0$ behaviour seen for H || c and at fields greater than 7 T, where a gradual decrease in 'n' is clearly seen with increase in Co concentration. These results point out that the currently available flux pinning models are inadequate to understand the behaviour of field exponent of $U_0$ and models that include the effect of disorder on bulk flux pinning exponent need to be formulated.

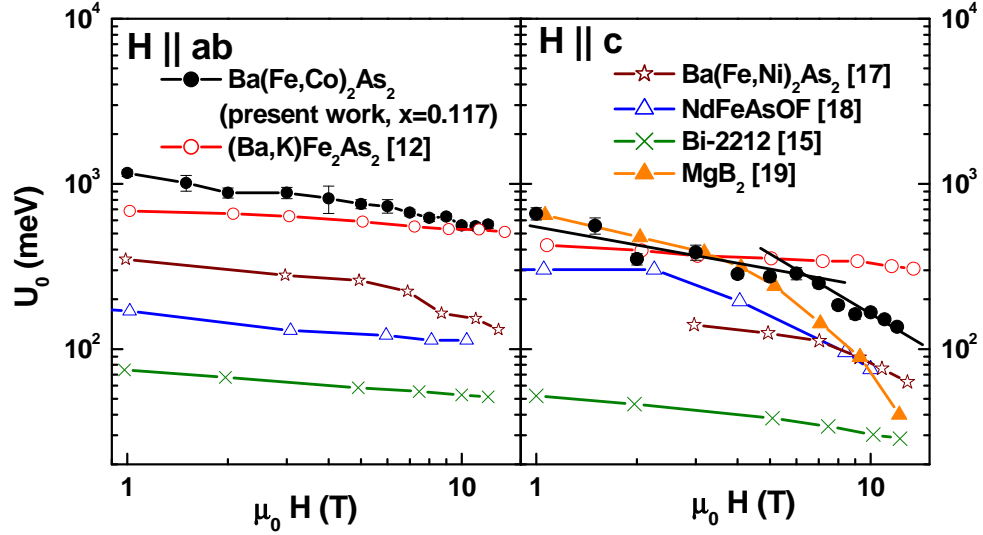

FIG 3: (Color online) Field dependence of $U_0$ for different types of superconductors compared with optimum doped $Ba(Fe,Co)_2As_2$ sample. Data for other superconductors are taken from literature.[12,15,17-19]

A comparison of the $U_0$ (H) obtained from the Co doped Ba122 samples with that observed in other FeAs based superconductors is shown in Fig. 3. It is evident from the Fig. 3a that $U_0$ in the optimally doped sample, viz., x=0.117 from the present set of measurements has the highest pinning potential for H || ab direction. The figure also suggests that other Ba122 superconductors viz., $(Ba,K)Fe_2As_2$[12] and $Ba(Fe,Ni)_2As_2$[17] have large $U_0$ which are weakly dependent on field. In comparison the $U_0$ values of the oxy-arsenide are smaller. The $U_0$ variation for H || c for the $MgB_2$ thin film is strongly dependent on field, although the low field $U_0$ value is comparable to that in the Ba122

compounds. In the HTS compound Bi2122, on the other hand the $U_0$ values are much smaller ascompared to that seen in the Ba122 compounds.

From the present study and flux imaging studies, it is compelling to suggest the large intrinsic pinning seen in the Co doped Ba122 arises due to the strong disorder induced by Co atoms. Recent band structure calculations [20] have suggested that all the extra electrons from the Co atom in the Fe sublattice tend to be confined to its muffin tin radius, resulting in a strong disorder around the dopant atom. The Co atom also induces an additional moment of 0.5 $\mu_B$/cell creating magnetic disorder, which due to magnetic exchange interaction with neighbouring Fe spreads to sizes larger than four unit cells.[21] The latter could result in the alteration of the superconducting order parameter around the substituting Co atom and being magnetic in nature could give rise to strong pinning. This is consistent with recent bitter decoration studies, which suggest that strong flux pinning could arise due of nano-scale spatial in-homogeneities in the superfluid density intrinsically, present in all arsenide superconductors.[22]

In summary, the flux pinning potential of Co doped $BaFe_2As_2$ superconductor has been investigated form magneto-resistance measurements. The $Ba(Fe,Co)_2As_2$ superconductor has very high intrinsic pinning potential that is field independent. This could result in large field independent $J_C$ observed in these systems. Further systematic investigations of the pinning potentials for other substitutions in $BaFe_2As_2$, correlated with local electron/spin density calculations may help clarify the intrigue of high intrinsic pinning strength observed in the Ba122 arsenide superconductors.

Vinod K acknowledges Department of Atomic Energy (DAE), India for the K.S. Krishnan Research Associateship.